\documentclass[preprint,aps,prb,onecolumn,superscriptaddress,floatfix,longbibliography,11pt]{revtex4-2}

\usepackage{amsmath,amssymb} 
\usepackage{bm} 
\usepackage{graphicx} 
\usepackage{textcomp} 
\usepackage[nolist,nohyperlinks]{acronym} 
\usepackage{float}
\usepackage{upgreek}
\usepackage{mdframed}
\usepackage{xcolor}
\usepackage{setspace}
\usepackage{siunitx} 
\usepackage{xr} 
\usepackage{enumitem} 
\usepackage{hyperref} 
\hypersetup{
    colorlinks=true,
    urlcolor=blue,
    citecolor=blue,
    linkcolor=blue,
}

\newcommand{\mytitle}{A swept-source dual-comb spectrometer on a chip}

\begin{document}

\title{\mytitle}

\author{Johannes Fuchsberger}
\affiliation{\textbf{These authors contributed equally to this work.}}
\affiliation{Institute of Solid State Electronics, TU Wien, 1040 Vienna, Austria}
\affiliation{Harvard John A. Paulson School of Engineering and Applied Sciences, Harvard University, Cambridge, MA 02138, USA}

\author{Theodore P. Letsou}
\email{tletsou@g.harvard.edu}
\affiliation{\textbf{These authors contributed equally to this work.}}
\affiliation{Harvard John A. Paulson School of Engineering and Applied Sciences, Harvard University, Cambridge, MA 02138, USA}

\author{Paul Chevalier}
\affiliation{Harvard John A. Paulson School of Engineering and Applied Sciences, Harvard University, Cambridge, MA 02138, USA}

\author{Marcus Ossiander}
\affiliation{Harvard John A. Paulson School of Engineering and Applied Sciences, Harvard University, Cambridge, MA 02138, USA}
\affiliation{Institute of Experimental Physics, Graz University of Technology, 8010 Graz, Austria}

\author{Dmitry Kazakov}
\affiliation{Harvard John A. Paulson School of Engineering and Applied Sciences, Harvard University, Cambridge, MA 02138, USA}

\author{Federico Capasso}
\email{capasso@seas.harvard.edu}
\affiliation{Harvard John A. Paulson School of Engineering and Applied Sciences, Harvard University, Cambridge, MA 02138, USA}

\author{Benedikt Schwarz}
\email{benedikt.schwarz@tuwien.ac.at}
\affiliation{Institute of Solid State Electronics, TU Wien, 1040 Vienna, Austria}
\affiliation{Harvard John A. Paulson School of Engineering and Applied Sciences, Harvard University, Cambridge, MA 02138, USA}

\date{\today}

\begin{abstract}
\textbf{Dual-comb spectroscopy (DCS) enables high-speed, high-resolution spectroscopy by down-converting optical spectra to the radio-frequency (RF) domain through the interference of two mutually detuned frequency combs.
Accurately resolving narrow molecular absorption features requires closely spaced comb lines generated by large laser cavities with long round-trip times, thereby hindering the miniaturization of high-resolution spectrometers.
Here, we circumvent this limitation by demonstrating a swept-source dual-comb spectrometer using two unidirectional racetrack semiconductor lasers integrated on the same chip.
The traveling-wave frequency combs are generated by strong RF injection at the laser cavity round-trip frequency of 14.3~GHz---otherwise known as quantum walk combs---and are continuously tuned in frequency by varying the laser drive currents.
This enables continuous spectral sampling over a 32~cm$^{-1}$ range centered near 8~\textmu m, with an effective sampling interval of approximately 80~MHz.
The outputs of both combs are coupled into the same monolithically integrated light coupler, providing automatic collinear alignment before the combined beam interrogates the sample.
We benchmark the spectrometer against an external-cavity tunable laser and validate its performance using HITRAN simulations for 1.25\% nitrous oxide.
Finally, the unidirectional comb architecture suppresses the detrimental effects of optical feedback, yielding comparable residuals even under deliberately introduced strong feedback.
More broadly, the benefits of miniaturization extend beyond reduced footprint: chip-scale integration enables new forms of electrical and optical control that can fundamentally change how dual-comb spectrometers are operated.}
\end{abstract}

\maketitle

\section{\label{sec:Intro}Introduction}

Laser frequency combs are bright light sources consisting of many coherent, precisely spaced optical frequencies. 
This regular frequency structure provides a precise ruler for optical measurements, enabling applications in precision spectroscopy, optical clocks, ranging, and frequency metrology \cite{Fortier2019, Hnsch2006, Hall2006}.
The highly ordered spectra of laser frequency combs, in which adjacent lines are separated by a common repetition frequency, emerge from a balance between gain, loss, dispersion, and nonlinear interactions within the laser.
By controlling this balance, frequency combs have been realized in a variety of platforms---from tabletop Ti:sapphire mode-locked lasers \cite{Spence1991} and fiber combs \cite{Fermann2013} to chip-scale sources such as Kerr microresonators \cite{Kippenberg2011} and semiconductor lasers \cite{Hugi2012}.
More recently, the development of chip-scale comb sources has become an important research direction for translating frequency-comb technology into compact, robust, and field-deployable systems. 
Semiconductor frequency combs are particularly attractive in this context because miniaturization provides more than just a reduction in footprint.
Direct electrical addressing of the laser provides key control knobs over the combs' output, which are difficult to realize with conventional bulk or fiber-based systems.

These advantages of integration are especially valuable in spectroscopic techniques, such as dual-comb spectroscopy (DCS), where control over the comb frequencies can directly determine how an optical spectrum is sampled.
DCS enables broadband optical spectra to be measured rapidly and with high resolution, without moving parts \cite{Coddington2016, Picqu2026}. 
It combines the broad spectral coverage of Fourier-transform infrared spectroscopy with the precision and coherence of optical frequency combs and has been demonstrated across the electromagnetic spectrum, from the ultraviolet \cite{Muraviev2024, Frst2024} to the terahertz \cite{Sterczewski2019, Yang2016}.
This versatility has led dual-comb spectrometers to be used in applications ranging from trace-gas monitoring \cite{Giorgetta2021} and molecular fingerprinting \cite{Timmers2018} to transient chemical dynamics \cite{Abbas2019}.

In DCS, two combs with slightly different repetition rates map optical-frequency differences onto the radio-frequency (RF) domain.
The combs have repetition rates $f_{\mathrm{rep}}$ and $f_{\mathrm{rep}}+\Delta$ (where $\Delta\ll f_{\mathrm{rep}}$), such that their optical tooth frequencies are
$f_1(n)=f_{\mathrm{CEO},1}+nf_{\mathrm{rep}}$
and
$f_2(n)=f_{\mathrm{CEO},2}+n(f_{\mathrm{rep}}+\Delta)$,
where $n$ is the comb tooth index and $f_{\mathrm{CEO},1}$ and $f_{\mathrm{CEO},2}$ are the carrier-envelope offset (CEO) frequencies. 
When the two combs interfere on a photodetector, corresponding pairs of optical comb teeth generate an RF comb with teeth spacing $\Delta$ that is a down-converted representation of the optical spectrum.
If one or both combs interrogate an absorbing sample, its transmission spectrum can be recovered by Fourier transforming the time-domain RF signal and normalizing it to a reference measurement acquired without the sample.

The native spectral sampling in DCS is set by the spacing between adjacent comb teeth, so features narrower than the comb spacing are undersampled.
High-resolution DCS has therefore traditionally relied on low-repetition-rate combs generated in long laser cavities \cite{Coddington2008, Coddington2010, Ideguchi2014}. 
This requirement presents a fundamental challenge for chip-scale integration, where short cavities naturally produce much larger comb spacings and therefore sparse spectral sampling.

Semiconductor laser frequency combs provide a natural way to circumvent this tradeoff while offering unusually direct electrical access to the parameters that define the comb spectrum, $f_{\mathrm{CEO}}$ and $f_{\mathrm{rep}}$. 
The DC injection current can continuously tune the optical comb offset without mode hops, while the small device capacitance enables efficient RF modulation superimposed on the DC bias for comb generation and stabilization of $f_{\mathrm{rep}}$. 
Semiconductor sources such as quantum cascade lasers (QCLs) and interband cascade lasers (ICLs) also emit directly in the mid-infrared molecular fingerprint region ($\lambda \sim 3$--12~\textmu m), avoiding the nonlinear frequency-conversion stages typically required by tabletop combs \cite{Qin2018, Lind2020, Ycas2018}. 
These properties make QCL and ICL frequency combs \cite{Hugi2012, Roy2024, Schwarz2019} particularly well suited for compact dual-comb spectroscopy \cite{Villares2014, MontesinosBallester2026}, since the comb spectrum can be swept continuously to fill the spectral gaps associated with the large mode spacing of short cavities \cite{Komagata2023, Gianella2020}. 
In principle, this enables spectrometer resolutions that approach the linewidth of the individual comb teeth rather than being limited by the native comb spacing. 
Moreover, both electrically driven combs required for DCS can be integrated together using on-chip light routing \cite{Letsou2025, MontesinosBallester2026}, simplifying optical alignment, improving mechanical stability, and reducing the overall system footprint.

In this work, we realize a swept-source dual-comb spectrometer based on two unidirectional racetrack semiconductor lasers integrated on a single millimeter-scale chip. 
Resonant RF injection at the cavity round-trip frequency converts the initially single-mode emission of each laser, centered near 8~\textmu m, into stabilized broadband frequency combs with milliwatt-level output powers at room temperature---otherwise known as quantum walk combs \cite{Heckelmann2023}. 
Such combs have been demonstrated across a variety of different laser platforms, such as dual-waveguide QCLs \cite{Cargioli2025}, terahertz lasers \cite{thz_qwc}, and interband diode lasers \cite{Marzban2026}, with applications spanning from spectroscopy \cite{MontesinosBallester2026, Heckelmann2026} to communication \cite{Marzban2026}. 
In case of dual comb spectroscopy, the RF drive delivered to each laser sets the repetition rates of the two combs and therefore their detuning, $\Delta$, while the DC injection currents tune the absolute optical frequencies of the comb teeth. 
By tuning the DC injection currents, we sweep both comb spectra over one free spectral range (FSR), filling the gaps between adjacent comb teeth and achieving continuous spectral coverage despite the large mode spacing of the compact cavities \cite{Komagata2023, Gianella2020}.
The outputs of the two racetrack lasers are combined through a monolithically integrated double adiabatic coupler into a common bus waveguide, providing intrinsically aligned outputs without external beam-combining optics. 
We validate the dual-comb spectrometer by measuring the transmission of bulk optical filters and 1.25\% nitrous oxide over a 32~cm$^{-1}$ spectral bandwidth, obtaining excellent agreement with measurements from an external-cavity tunable laser and with HITRAN simulations. 
Finally, the unidirectional racetrack architecture provides intrinsic robustness against optical feedback: comparable spectral residuals are maintained even when 50\% of the emitted light is reflected back toward the laser cavity, eliminating the need for bulky, high-insertion-loss optical isolators.

\section{\label{sec:2}The on-chip dual-comb spectrometer}

\begin{figure*}[t]
    \centering
    \includegraphics[clip=true,width=\textwidth]{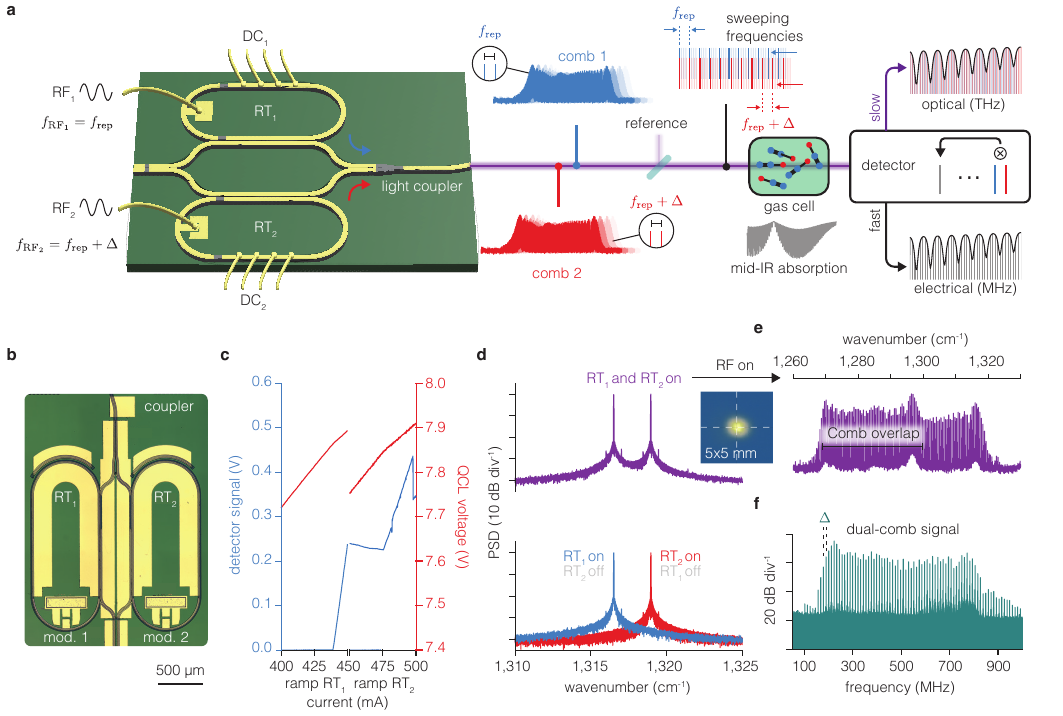}
    \caption{\textbf{Overview of the on-chip swept-source dual-comb spectrometer.}
    \textbf{a,} Operating principle. Two unidirectional racetrack quantum cascade lasers, RT$_1$ and RT$_2$, are integrated on the same chip and driven independently by DC and RF signals. The DC biases bring the racetracks above threshold and tune their optical frequencies, while resonant RF injection at $f_{\mathrm{RF},1}=f_{\mathrm{rep}}$ and $f_{\mathrm{RF},2}=f_{\mathrm{rep}}+\Delta$ generates and stabilizes two frequency combs with repetition-rate detuning $\Delta$. The comb outputs are combined on chip by a double adiabatic coupler into a common bus waveguide. A small fraction of the combined light is sampled as a reference, while the remainder interrogates the sample. Sweeping the DC biases translates the optical comb spectra between adjacent comb teeth, enabling continuous spectral sampling despite the large native comb spacing. After photodetection, pairs of optical comb teeth beat to form an RF comb with tooth spacing $\Delta$, down-converting the optical spectrum from the THz to the MHz domain for rapid electrical acquisition.
    \textbf{b,} Microscope image of the fabricated dual-racetrack chip, showing the two racetrack cavities (RT), RF injection sections (mod), DC contacts, bus waveguides, and output coupler. Each RF section is contacted through ground--signal--ground (GSG) pads, which can be accessed either by landing an RF probe directly on the chip or by wire bonding the pads to an external RF interface. Each racetrack has a repetition rate of approximately 14.3~GHz, corresponding to a cavity perimeter of 6.1~mm. The ridge waveguides are dry etched to a width of 10~\textmu m with nearly vertical sidewalls.
    \textbf{c,} Light--current--voltage characteristics of the two independently addressable racetrack lasers. RT$_1$ is ramped from 400 to 450~mA and RT$_2$ from 450 to 500~mA. The ``kinks'' in the voltage traces mark the onset of lasing, while the detector signal records the emitted optical power. The sharp decrease in detector signal near 500~mA indicates a mode hop.
    \textbf{d,} Power spectral densities (PSDs) and far-field beam profile demonstrating independent operation and spatial overlap of the two racetracks. With both racetracks biased, the spectrum contains the two single-mode lasing peaks of RT$_1$ and RT$_2$ with minimal frequency pulling. The inset shows the measured output beam profile with both racetracks operating. The image size is 5~mm $\times$ 5~mm.
    \textbf{e,} Broadband optical spectra generated when resonant RF injection is applied to both racetracks. The shaded region marks the spectral overlap between the two combs and therefore the optical bandwidth available for dual-comb detection.
    \textbf{f,} RF spectrum of the multiheterodyne dual-comb signal produced by beating the overlapping optical comb teeth on a fast detector. The RF comb tooth spacing is the repetition-rate detuning $\Delta$ between the two optical combs.}
    \label{fig:1}
\end{figure*}

Figure~\ref{fig:1}\textbf{a} shows a schematic of our on-chip dual-comb spectrometer, which is based on QCL material emitting near a wavelength of 8~\textmu m. 
The device consists of two traveling-wave racetrack resonators, RT$_1$ and RT$_2$, each independently driven by DC and RF signals, along with two bus waveguides separated from the racetracks by a 1~\textmu m air gap, and an integrated light coupler \cite{Kazakov2024, Kacmoli2022}.
The two racetrack resonators act as our comb generators for the spectrometer.
Light generated in each racetrack via electrical pumping is out-coupled into its corresponding bus waveguide and routed to the light coupler. 
The coupler is formed from a double adiabatic waveguide taper, which evanescently transfers light from both bus waveguides into a single output waveguide (see Supplementary Information Section 1). 
This architecture combines the emission from the two racetracks into a collinear beam directly on chip, eliminating the need for off-chip beam combining. 
The RF sections are implemented as ground-signal-ground (GSG) contacts along the curved sections of the racetracks, which support either direct injection with GSG probes or wire-bonding to an RF transmission line (see Supplementary Information Section 2). 
The parasitic capacitance of the RF contacts is minimized by depositing a 1.5~\textmu m insulating layer beneath the metal contacts. 
As a result, the RF modulation bandwidth exceeds 10~GHz, enabling RF injection near the cavity round-trip frequency of 14.33~GHz \cite{Letsou2026}.

Frequency-comb operation is achieved by applying a resonant RF tone on top of the DC drive current. 
The resonant RF injection modulates the intracavity gain at the cavity round-trip frequency and couples adjacent longitudinal modes into a frequency comb.
In an actively modulated laser, the steady-state solutions in the frequency domain can be described by Hermite--Gauss spectral envelopes, with the lowest-order solution corresponding to the Gaussian pulse---commonly associated with conventional active mode locking \cite{Haus1975}. 
In QCLs, however, the sub-picosecond gain recovery time suppresses the formation of isolated pulses and stabilizes the comb spectral envelope to a superposition of higher-order Hermite--Gauss modes \cite{Heckelmann2023}.
These states are referred to as quantum walk combs as the ballistic expansion of the laser modes can be described using a quantum random walk \cite{Dikopoltsev2025}.
This makes modulated QCLs especially attractive for DCS because their output spectra are broadband, relatively flat, and repetition-rate stabilized, while avoiding high-peak-power pulses that could induce a nonlinear absorption response of the sample.

The chip operates as a dual-comb spectrometer by driving the two racetrack lasers with RF frequencies separated by a small detuning, $\Delta$. 
The two comb outputs are combined on chip and sent through the sample, while a small fraction is split off as a reference. 
After photodetection, pairs of optical comb teeth beat together to produce a radio-frequency comb with spacing $\Delta$, down-converting the optical spectrum from the THz to the MHz domain. 
The DC currents tune the optical comb frequencies, allowing the combs to be swept between adjacent teeth for continuous spectral sampling.
This enables rapid acquisition of a high-resolution absorption spectrum of the gas sample without the slow mechanical scanning required by conventional spectrometers.

A microscope image of the chip is shown in Fig.~\ref{fig:1}\textbf{b}, where each independently addressable section is defined by separate gold contacts. 
In practice, wire bonds, rather than probes, are used to connect to each DC and RF section of the chip (see Supplementary Information Section 3). 
The dry-etch process used to fabricate the chip forms 8~\textmu m-deep trenches between the different device sections, providing excellent electrical isolation between different laser elements. 
This isolation is demonstrated in Fig.~\ref{fig:1}\textbf{c}, which shows the light-current-voltage (LIV) characteristics of the two racetracks during sequential biasing using a single photodetector placed at the facet of the coupler. 
First, RT$_1$ is biased above threshold by sweeping its current from 400~mA to 450~mA.
At threshold, an optical signal appears on the photodetector.
RT$_1$ is then held above threshold while the current applied to RT$_2$ is swept from 450~mA to 500~mA, bringing RT$_2$ above threshold as well. 
Once RT$_2$ begins lasing, light from both racetracks is combined into the same waveguide. 
The measured signal on the detector therefore increases as the second racetrack contributes additional optical power at the coupler facet.

Despite the use of multiple current sources to bias the chip, the output spectrum of each racetrack remains nearly unaffected by the biasing condition of the other \cite{Fuchsberger2025}, as shown in Fig.~\ref{fig:1}\textbf{d}. 
In the absence of RF injection, each racetrack laser typically emits a single longitudinal mode as shown via their power spectral densities (PSDs) measured using a home-built Fourier transform infrared spectrometer (FTIR).
When both racetracks are biased simultaneously, the two laser colors are combined without significant frequency pulling or spectral distortion. 
Moreover, when collimated with a lens, the combined beam from both racetracks exhibits excellent spatial quality, with Gaussian full width at half maximum values of 1.19~mm and 1.00~mm along the $x$ and $y$ directions, respectively, measured 0.5~m from the chip.

Applying resonant RF injection at a power of 30~dBm to the lasers broadens their single-mode spectra into overlapping frequency combs, as shown in Fig.~\ref{fig:1}\textbf{e}; direct injection through the use of GSG probes broadens the combs further, indicating non-negligible RF losses through the transmission line and wire bonds (see Supplementary Information Section 4). 
The resulting dual-comb signal in the RF domain, produced by multiheterodyne beating between adjacent comb teeth, is shown in Fig.~\ref{fig:1}\textbf{f}. 
The RF signal spans from approximately 100~MHz to over 900~MHz, occupying the full bandwidth of the mercury cadmium telluride (MCT) detector.

\section{\label{sec:3}Acquiring the dual-comb trace}

One of the major advantages of our on-chip dual-comb spectrometer is the simplicity of the signal correction required for coherent averaging. 
In a typical free-running dual-comb spectrometer, fluctuations in both $f_{\mathrm{CEO}}$ and $f_{\mathrm{rep}}$ cause light from both combs to arrive at the detector with different timing and phase, preventing direct coherent averaging without signal correction \cite{Burghoff2016, Sterczewski2019_2, Burghoff2019}. 
In our system, however, the repetition rates of both combs are stabilized by RF injection, leaving only their relative $f_{\mathrm{CEO}}$ drift to correct. 
We compensate for this drift by tracking the phase of a single RF comb tooth using a Hilbert-transform-based algorithm \cite{Sterczewski2019_2}, since the phase evolution of a single RF line directly captures the relative CEO drift. 
Because only a single phase correction is required, the method is computationally simple and well suited to real-time implementation \cite{Eber2025, Walsh2024}.

\begin{figure*}[t]
    \centering
    \includegraphics[clip=true,width=\textwidth]{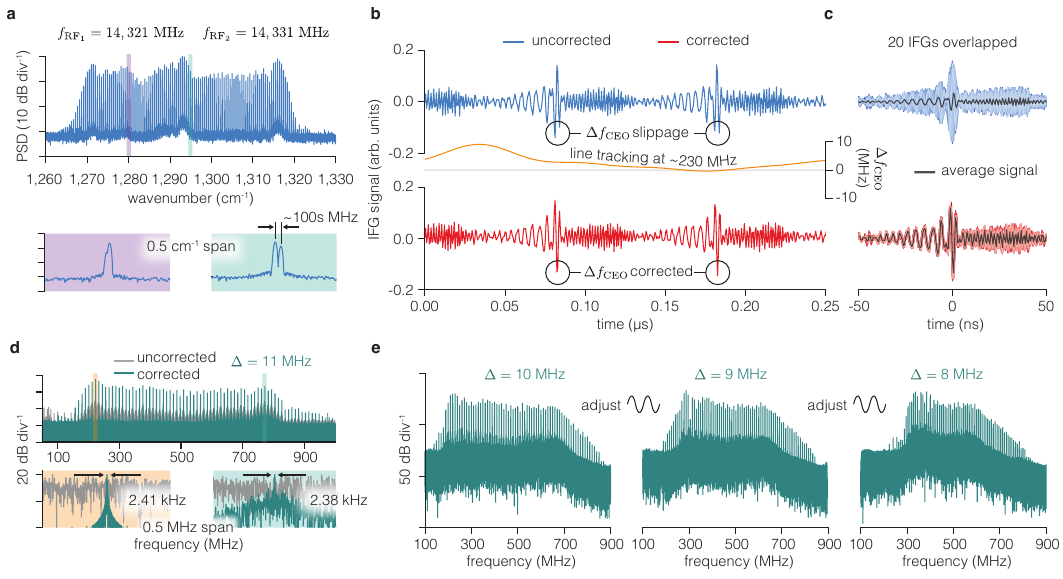}
    \caption{\textbf{Controlling the dual-comb spectrometer.}
    \textbf{a,} High-resolution optical spectrum of the dual-comb spectrometer operating with repetition rates $f_{\mathrm{RF},1}=14.321$~GHz and $f_{\mathrm{RF},2}=14.331$~GHz, corresponding to a repetition-rate detuning of $\Delta = 10$~MHz. The purple and blue shaded regions are expanded below, showing the relative walk-off between the two combs. The maximum tooth-to-tooth frequency separation is on the order of hundreds of MHz.
    \textbf{b,} Raw and corrected time-domain RF beat signals measured on a fast detector. Interferogram (IFG) bursts occur every $1/\Delta$, corresponding to 100~ns for $\Delta = 10$~MHz. RF injection stabilizes the repetition-rate detuning, while residual phase jitter in the uncorrected trace arises because the CEO frequencies of the two combs are not stabilized. This produces a time-dependent $\Delta f_{\mathrm{CEO}}$, shown in the inset as extracted by tracking a dual-comb tooth near 230~MHz. After tracking, the CEO phase is corrected and shifted to zero frequency, yielding the corrected time trace.
    \textbf{c,} Coherent averaging of 20 sequential IFGs before and after CEO correction. Without correction, phase jitter prevents coherent buildup of the averaged signal. After correction, the IFGs align in phase and can be coherently averaged to improve the signal-to-noise ratio.
    \textbf{d,} Effect of CEO correction for $\Delta = 11$~MHz. In the uncorrected RF spectrum, individual comb lines are broadened and largely buried in the noise floor. After correction, the RF comb lines narrow to the kHz level, as shown in the expanded spectral regions.
    \textbf{e,} Tunability of the repetition-rate detuning. Adjusting $\Delta$ from 10 to 9 and 8~MHz changes the mapping of the optical comb spectrum into the RF domain, thereby tuning the RF bandwidth occupied by the dual-comb signal.}
    \label{fig:2}
\end{figure*}

Figure~\ref{fig:2}\textbf{a} shows the optical spectrum of the dual-comb laser for a repetition-rate detuning of $\Delta=10$~MHz, measured with an FTIR. 
At this small detuning, the comb teeth from the two racetracks are nearly indistinguishable in the optical spectrum except near the edges of the overlap region, where the relative frequency offset between corresponding teeth becomes large enough to exceed the FTIR resolution. 

For dual-comb detection, the combined optical output is instead sent directly onto a fast photodetector. 
The detector converts the multiheterodyne beating between the two optical combs into a time-domain electrical signal consisting of a periodic train of interferogram (IFG) bursts. 
For $\Delta=10$~MHz, successive bursts are separated by $1/\Delta=100$~ns. 
The corresponding uncorrected detector trace is shown in Fig.~\ref{fig:2}\textbf{b} (blue). 
Although the RF injection stabilizes the repetition rates of the two combs, residual relative drift in $f_{\mathrm{CEO}}$ causes the phase of successive IFGs to vary in time and prevents direct coherent averaging. 
Applying the drift correction described above produces the corrected IFG shown in red, while the yellow trace in the inset shows the extracted $\Delta f_{\mathrm{CEO}}$ used for the computational phase correction.
Once corrected, successive IFGs can be coherently averaged to improve the signal-to-noise ratio. 
Figure~\ref{fig:2}\textbf{c} compares the average of twenty sequential IFGs before and after correction, demonstrating the substantial improvement in interferogram contrast and coherence.

Applying the same phase-correction procedure to data acquired at a repetition-rate detuning of $\Delta=11$~MHz yields the RF spectra shown in Fig.~\ref{fig:2}\textbf{d}. 
Before correction, fluctuations in the relative comb phase broaden the RF beat notes and obscure the individual comb teeth. 
After correction, the RF comb teeth collapse to linewidths on the order of kilohertz across the measured bandwidth, demonstrating that the mutual coherence of the two combs can be recovered computationally.
Because the repetition rates of the two racetracks are set by the applied RF drives, the detuning $\Delta$---and therefore the spacing between adjacent RF comb teeth---can be selected electronically. 
In our system, stable dual-comb operation is obtained for detunings between 8 and 11~MHz. 
Representative corrected RF spectra for $\Delta=8$, 9, and 10~MHz are shown in Fig.~\ref{fig:2}\textbf{e}. 
Varying $\Delta$ changes the total multiheterodyne bandwidth, allowing the RF spectrum to be tailored to the available detector and acquisition bandwidth.

\section{\label{sec:4}Benchmarking the dual-comb spectrometer}

\begin{figure*}[t]
    \centering
    \includegraphics[clip=true,width=\textwidth]{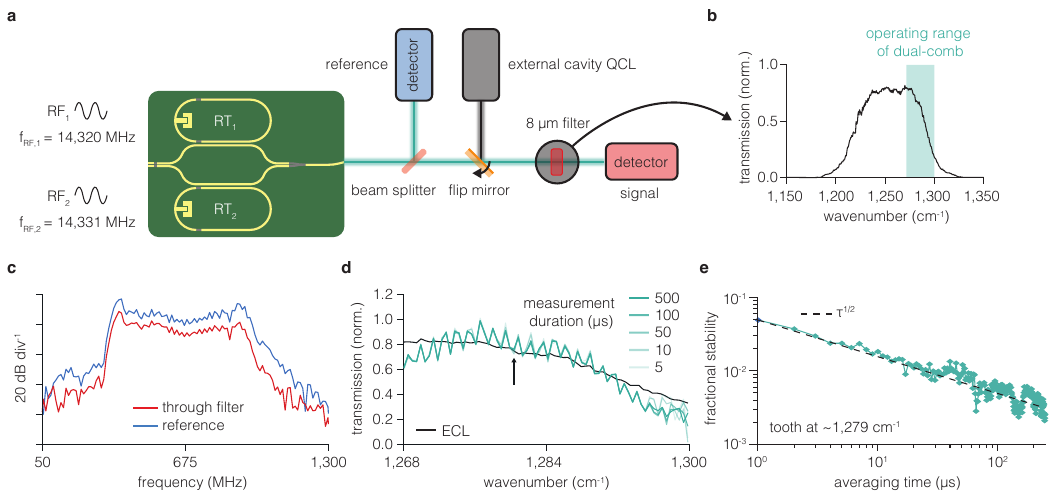}
    \caption{\textbf{Benchmarking the dual-comb spectrometer with an 8~\textmu m bandpass filter.}
    \textbf{a,} Measurement configuration. Two racetrack lasers driven at $f_{\mathrm{RF},1}=14.320$~GHz and $f_{\mathrm{RF},2}=14.331$~GHz generate frequency combs with a repetition-rate detuning of $\Delta=11$~MHz. The combs are combined on chip and directed through a beam splitter. One arm provides a reference measurement of the incident dual-comb spectrum, while the second passes through the 8~\textmu m bandpass filter before detection. A flip mirror directs an external-cavity tunable laser (ECL) through the same filter for an independent transmission measurement.
    \textbf{b,} Bandpass-filter transmission measured with the ECL. The shaded region indicates the optical frequency range sampled by the dual-comb spectrometer.
    \textbf{c,} Reference and filtered RF spectral envelopes obtained from 500~\textmu s detector time traces. Each time trace is divided into intervals corresponding to individual IFG periods, Fourier transformed, and averaged. Attenuation of the filtered spectrum toward higher RF frequencies corresponds to the falling edge of the optical filter passband.
    \textbf{d,} Filter transmission recovered from the dual-comb measurement for acquisition times from 5 to 500~\textmu s. The transmission is obtained by dividing the filtered spectrum by the reference spectrum, mapping the RF frequency axis to optical wavenumber, and scaling the result to the ECL reference. Even the 5~\textmu s measurement reproduces the overall ECL response, while the finer dual-comb sampling resolves etalon fringes from the approximately 2-mm-thick filter that are not resolved by the ECL scan.
    \textbf{e,} Fractional stability of a representative spectral tooth near 1,279~cm$^{-1}$ as a function of averaging time. The stability follows approximately $\tau^{-1/2}$ over the measured range, consistent with averaging of predominantly uncorrelated noise.}
    \label{fig:3}
\end{figure*}

Having established stable dual-comb operation, controllable repetition-rate detuning, and coherent averaging of the interferograms, we next benchmark the spectrometer against an independent optical reference.
We use an 8~\textmu m bandpass filter with a 0.5~\textmu m passband as a test sample using the setup described in Fig.~\ref{fig:3}\textbf{a}. 
The filter transmission is first characterized using an external-cavity tunable laser (ECL), whose measured response is shown in Fig.~\ref{fig:3}\textbf{b}. 
The ECL provides a reference measurement over the operating range of the dual-comb spectrometer, indicated by the green shaded region. 
Owing to mode hops introduced by the external cavity, the ECL scan has an effective spectral resolution of approximately 50~GHz, limiting its ability to resolve fine spectral features in the filter \cite{Wysocki2008}.

The same filter is then measured using the on-chip dual-comb spectrometer. 
For both the reference and sample measurements, we record a 500~\textmu s detector time trace containing many successive IFG bursts. 
The trace is divided into intervals corresponding to a single IFG period, and each interval is Fourier transformed to obtain an RF spectrum. 
Averaging these spectra produces the reference and sample envelopes shown in Fig.~\ref{fig:3}\textbf{c}. 
The filter response is directly visible in the RF domain. 
Compared with the reference, the sample spectrum is increasingly attenuated toward the high-frequency edge, corresponding to the falling edge of the optical passband.
The filter transmission is obtained by dividing the sample spectrum by the reference spectrum, converting the RF frequency axis to optical frequency, and scaling the result to the ECL reference. 
Figure~\ref{fig:3}\textbf{d} shows the resulting dual-comb transmission for several acquisition times together with the ECL measurement.
Even for acquisition times as short as 5~\textmu s, the dual-comb spectrometer accurately reproduces the overall filter transmission measured by the ECL. 
At the same time, the finer spectral sampling of the dual-comb measurement reveals oscillations across the passband that are not resolved in the ECL trace. 
These oscillations arise from etalon fringes in the approximately 2~mm-thick silicon optic \cite{ThorlabsFB8000500}. 
Near the edges of the measured spectral window, the dual-comb transmission deviates from the ECL reference because the decreasing comb power lowers the signal-to-noise ratio of the corresponding RF teeth.

The measurement stability is quantified in Fig.~\ref{fig:3}\textbf{e}, which shows the fractional stability of a representative comb tooth near 1,279~cm$^{-1}$. 
The stability improves approximately as $\tau^{-1/2}$ with averaging time, indicating that the measurement is dominated by uncorrelated noise over the measured averaging interval.

\section{\label{sec:5}Continuous frequency sweeping of the dual-comb spectrometer}

Many molecular absorption features in the mid-infrared consist of narrow ro-vibrational transitions with linewidths below 500~MHz at atmospheric pressure. 
Resolving these features in our spectrometer therefore requires spectral sampling substantially finer than the spacing between adjacent comb teeth. 
This presents a challenge for chip-scale dual-comb spectrometers, whose compact cavities naturally produce repetition rates of tens of gigahertz and correspondingly sparse optical sampling grids \cite{Kippenberg2011, Kuse2020}. 
As a result, narrow absorption features can fall between neighboring comb teeth and remain undetected \cite{Yu2018, Suh2016}.

Our spectrometer overcomes this limitation by continuously sweeping the optical frequencies of both combs while maintaining dual-comb operation. 
Independent control of the laser drive currents and RF injection frequencies allows the comb teeth to be translated over approximately one FSR while preserving a fixed repetition-rate detuning, $\Delta$. 
Spectra acquired at successive tuning points are then stitched together to provide continuous spectral sampling across the entire emission bandwidth \cite{Komagata2023, Gianella2020}. 
The reproducibility of the electrical tuning allows successive sweeps to return to the same optical frequencies over multiple days without requiring a simultaneous external frequency reference.

\begin{figure*}[t]
    \centering
    \includegraphics[clip=true,width=\textwidth]{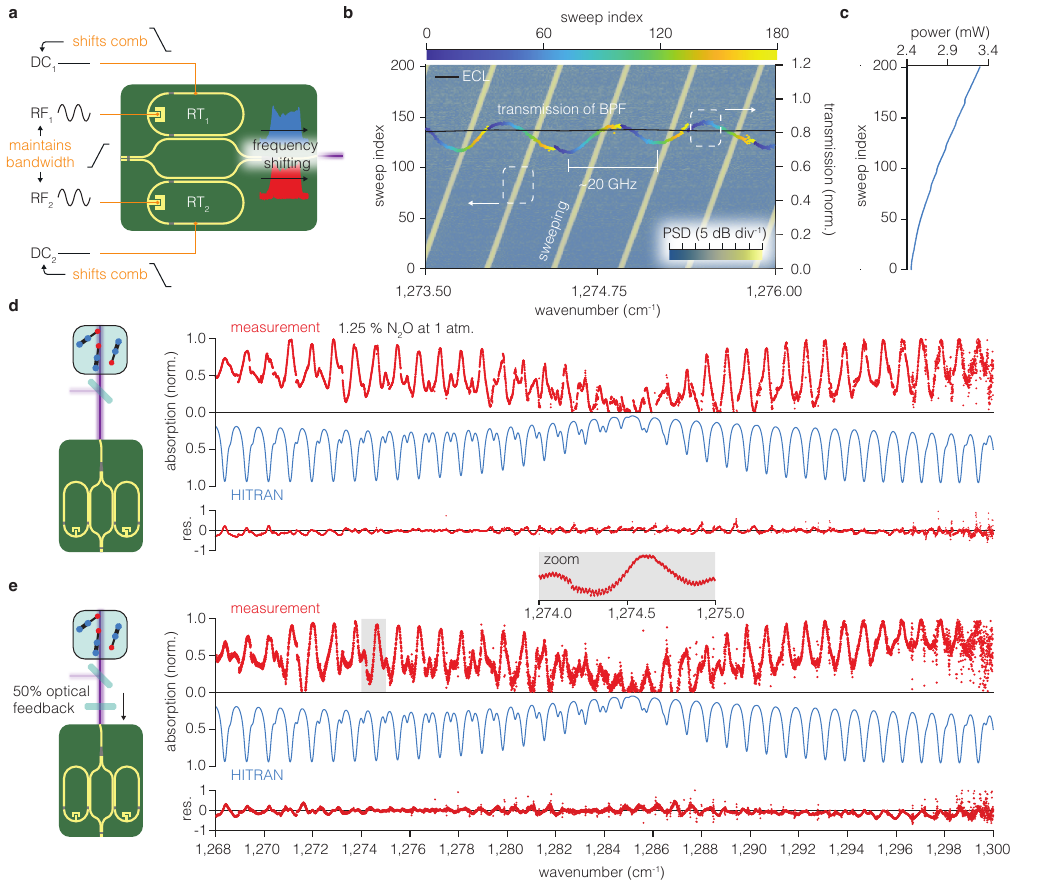}
    \caption{\textbf{Swept-source dual-comb spectroscopy and optical-feedback robustness.}
    \textbf{a,} Schematic of the frequency-sweeping scheme. The DC injection currents shift the optical frequencies of the two combs, while the RF injection frequencies are adjusted concurrently to maintain broadband comb operation and a fixed repetition-rate detuning, $\Delta$, throughout the sweep. In this way, the comb teeth are translated across the spectral gaps between their native frequencies while preserving dual-comb operation.
    \textbf{b,} Continuous optical-frequency sweep measured with an FTIR while the dual-comb spectrometer interrogates the 8~\textmu m bandpass filter. The background color map shows a portion of the PSD as a function of optical wavenumber and sweep index; the bright diagonal traces correspond to individual comb teeth swept continuously in frequency. Superimposed is the bandpass-filter transmission reconstructed from the dual-comb measurement, together with the independently measured ECL transmission. Approximately one FSR is covered in 180 sweep steps, corresponding to an effective spectral sampling interval of approximately 80~MHz. The finer sampling resolves the $\sim20$~GHz etalon fringes of the filter that are not resolved in the ECL scan.
    \textbf{c,} Total laser output power measured over the sweep. The power remains between approximately 2.4 and 3.4~mW across the 201 operating points.
    \textbf{d,} Swept dual-comb measurement of 1.25\% N$_2$O in air at 1~atm using a 13~cm-long single-pass gas cell with anti-reflection-coated ZnSe windows. The measured absorption spectrum is compared with a HITRAN-based reference over 1,268--1,300~cm$^{-1}$. The residuals remain small across the measured bandwidth.
    \textbf{e,} Measurement of the same N$_2$O sample with approximately 50\% optical feedback deliberately reflected toward the chip. Despite the strong feedback, the measured absorption remains in close agreement with the HITRAN reference and the residuals remain comparable to those obtained without deliberate feedback, demonstrating the intrinsic feedback robustness of the unidirectional racetrack architecture.}
    \label{fig:4}
\end{figure*}

Figure~\ref{fig:4}\textbf{a} illustrates the continuous frequency-sweeping scheme. 
The DC injection currents shift the optical frequencies of the two combs, while the RF injection frequencies are adjusted concurrently to maintain broadband comb operation and a fixed repetition-rate detuning, $\Delta$, throughout the sweep. 
In this way, the comb teeth are translated across the spectral gaps between their native frequencies without sacrificing dual-comb operation.

We first demonstrate continuous spectral sampling using the 8~\textmu m bandpass filter introduced in Fig.~\ref{fig:3}. 
Figure~\ref{fig:4}\textbf{b} shows the optical spectrum of the dual-comb laser measured by the FTIR as the two combs are swept through 201 operating points. 
Bright diagonal traces correspond to individual optical comb teeth shifting continuously in frequency as the drive currents are varied. 
Superimposed on the FTIR data is the filter transmission reconstructed from the dual-comb measurement at each sweep point, together with the independently measured ECL transmission for comparison. 
As the comb teeth sweep across the gaps between their native frequencies, the dual-comb spectrometer samples the filter response continuously rather than only at the discrete 14.33~GHz comb spacing. 
A total of 180 sweep points translates the comb spectrum by approximately one FSR, corresponding to an effective spectral sampling interval of approximately $14.33~\mathrm{GHz}/180 \approx 80~\mathrm{MHz}$. 
The reconstructed transmission follows the ECL response while also resolving the $\sim20$~GHz-period etalon fringes that are not captured by the more coarsely sampled ECL measurement. 
Because each dual-comb acquisition requires only 500~\textmu s, a complete 180-point sweep is acquired in approximately 90~ms. 
Figure~\ref{fig:4}\textbf{c} shows that the total output power remains between approximately 2.4 and 3.4~mW over the sweep.

We next apply the swept spectrometer to gas-phase absorption spectroscopy of nitrous oxide (N$_2$O), as shown in Fig.~\ref{fig:4}\textbf{d}. 
The sample consists of 1.25\% N$_2$O in air at atmospheric pressure contained in a 13~cm-long gas cell with anti-reflection-coated ZnSe windows and interrogated in a single-pass geometry. 
The stitched dual-comb spectrum spans 1,268--1,300~cm$^{-1}$ and is compared with a HITRAN-based reference spectrum \cite{HITRAN2024}, whose parameters are established from an independent FTIR measurement of the same gas cell (Supplementary Information Section~5). 
The measured spectrum reproduces the positions and relative strengths of the N$_2$O absorption lines across the full bandwidth, with small residuals relative to the HITRAN model. 

\section{\label{sec:6}Intrinsic robustness to optical feedback}

Optical feedback is one of the principal challenges limiting the deployment of compact dual-comb spectrometers \cite{ Hillbrand2018, Burghoff2015}. 
Reflections from downstream optics or the sample can re-enter the laser cavity, destabilizing the comb state through excess phase noise, linewidth broadening, or complete collapse to single-mode operation. 
Optical isolators are therefore commonly employed to suppress feedback \cite{Wen2026, Tian2022}, but occupy valuable space and, in the mid-infrared, can introduce insertion losses approaching 30\% \cite{ThorlabsI4500W4}.

The unidirectional racetrack lasers used here are intrinsically robust against optical feedback, despite possessing linewidth-enhancement factors ($\alpha$) comparable to those of other semiconductor lasers \cite{Opaak2021}, for which the effective feedback strength scales as $\sqrt{1+\alpha^2}$ \cite{Acket1984}. 
Back-reflected light re-enters the cavity counter-propagating to the primary lasing field, which strongly reduces its coupling to the circulating comb state, making the comb state largely insensitive to delayed optical feedback \cite{Opacak2024, Letsou2026}.

To demonstrate this robustness, we deliberately introduce strong optical feedback using a 50\% beamsplitter placed directly in front of the collimated output beam. 
The alignment is optimized by maximizing the photovoltage generated on the coupler contact, thereby maximizing the amount of reflected light coupled back into the device.
As shown in Fig.~\ref{fig:4}\textbf{e}, the measured N$_2$O absorption spectrum remains in excellent agreement with the HITRAN reference even under this strong-feedback condition. 
Although the feedback introduces weak oscillations into the dual-comb signal, shown in the inset, no mode hops or collapse of the comb state are observed. 
The residuals remain comparable to those obtained without deliberate feedback, demonstrating that the unidirectional racetrack architecture preserves stable dual-comb operation even in the presence of substantial back-reflections.

\section{\label{sec:7}Outlook}

Since its introduction in 2004, DCS has emerged as one of the most powerful techniques for broadband, high-resolution spectroscopy \cite{Keilmann2004}. 
Despite its remarkable progress, FTIR spectrometers continue to dominate the commercial infrared spectroscopy market. 
This naturally raises the question: where can chip-scale dual-comb spectrometers provide unique value?

In the mid-infrared, QCL-based dual-comb spectrometers are particularly well suited to applications requiring high temporal resolution, stand-off detection, or field-deployable operation. 
The microsecond acquisition times enabled by DCS make these systems attractive for studying transient phenomena, including gas-phase reaction kinetics, combustion diagnostics, and dynamical processes in proteins and other biomolecules \cite{Hayden2024}. 
Their small footprint also makes them promising candidates for stand-off chemical sensing, where the instrument must be brought to the sample rather than vice versa. 
Although portable Raman spectrometers already serve this role in many applications \cite{Misra2019}, their spectral resolution is often insufficient for resolving narrow gas-phase transitions. 
By contrast, a swept-source mid-infrared dual-comb spectrometer can provide high-resolution spectra at rapid acquisition rates, particularly when paired with real-time signal processing.
Real-world environments, however, are rarely compatible with laboratory-grade optical systems. 
Mechanical vibrations, temperature fluctuations, optical misalignment, and parasitic back-reflections can all degrade the performance of conventional dual-comb spectrometers. 
The monolithic beam-combining architecture and intrinsic feedback robustness of the unidirectional racetrack lasers directly address these challenges.

Several opportunities remain for further improving the demonstrated platform. 
Precise control over the relative optical frequencies of the two combs remains important, since only overlapping comb teeth contribute to the dual-comb signal. 
Recent demonstrations of QCL frequency combs spanning more than 100~cm$^{-1}$ suggest that future devices could simultaneously interrogate multiple molecular species across much broader spectral bandwidths \cite{Zeng2025}, enabling highly multiplexed gas sensing \cite{Tian2023}.

The present implementation is also limited to spectral resolutions broader than the intrinsic linewidth of the comb teeth, which for ring QCLs could be below 100~kHz \cite{Parriaux2026}. 
This limitation arises because both combs interrogate the sample, causing the measured absorption response to be convolved with the spectral profiles of two comb teeth. 
In principle, this response could be deconvolved provided the relative comb frequencies are known with sufficient accuracy. 
Alternatively, future architectures that route only one comb through the sample while using the second solely as a local oscillator could approach the intrinsic linewidth of the comb teeth while preserving the alignment stability afforded by monolithic integration.

More broadly, this work illustrates how photonic integration can improve functionality rather than simply reduce footprint. 
By reshaping how the combs are generated, controlled, and routed, integration enables an architecture that is more capable than a direct miniaturization of conventional dual-comb systems. 
We anticipate that continued advances in semiconductor frequency combs will establish swept-source dual-comb spectroscopy as a practical platform for molecular sensing throughout the mid-infrared.

\section*{Methods}

\textbf{Device design and fabrication.} 
The nominal racetrack design closely follows that of Ref.~\cite{Letsou2026}. 
The QCLs consist of alternating InGaAs/InAlAs layers epitaxially grown on an InP substrate. 
The active region is based on a single-phonon-continuum depopulation scheme and emits near a center wavelength of 7.9~\textmu m. 

The devices are defined using photolithography with Si$_3$N$_4$ as an etch mask. 
The 10~\textmu m-wide waveguides are dry-etched using Ar:Cl$_2$ chemistry to a depth of 8~\textmu m, yielding nearly vertical sidewalls. 
A 250~nm insulating layer of Si$_3$N$_4$ is then deposited and opened on the top contact of the laser ridge to enable electrical biasing. 
An additional 1250~nm layer of Si$_3$N$_4$ is deposited along the curved sections of both racetracks to reduce the parasitic capacitance of the signal contact in the modulation section. 
Ground planes are etched adjacent to the signal contact, enabling a GSG probe to land directly on the laser chip. 
The ground and signal contacts are also made sufficiently wide to support wire bonds, which are used to deliver RF signals in the experiments.

The curved sections of each racetrack have a radius of 500~\textmu m, while the straight sections are 1500~\textmu m long, resulting in an FSR of approximately 14.3~GHz. 
Both racetracks are coupled to bus waveguides through a 1~\textmu m air gap, which is achievable using optical lithography. 
Light exiting the racetracks through the bus waveguides is combined into a single output waveguide using a double adiabatic coupler.
Each bus waveguide and the combiner waveguide taper their widths in opposite directions over a length of 500~\textmu m while maintaining a 1~\textmu m air gap between adjacent sections.

The laser chips are cleaved to define the output facets of the coupler, indium-soldered epitaxial-side up onto copper submounts, and cooled to 16$^{\circ}$C using Peltier units driven by Wavelength Electronics TC15 temperature controllers. 
Light from the coupler is collimated using an aspheric lens from Thorlabs (C093TME-F).
Each section of the laser is wire-bonded to custom circuit boards for independent electrical driving using low-noise DC current supplies.
In these experiements, Wavelength Electronics QCL2000 Lab units are used for the racetrack sections, and Wavelength Electronics QCL1500 Lab units are used for the waveguide sections and the coupler. 
RF signals from a SynthHD PRO source from Windfreak Technologies are amplified using a Mini-Circuits ZVE-3W-183+ amplifier and fed into custom RF transmission lines optimized for low loss at frequencies beyond 10~GHz.

\textbf{Dual-comb measurements and reconstruction.} 
We first benchmark the dual-comb spectrometer using a commercial 8.0~\textmu m bandpass filter with a nominal full width at half maximum of 500~nm (Thorlabs FB8000-500). 
The filter is placed in the combined output path of the two racetrack QCLs and is used as a stable, broadband spectral feature for comparing the dual-comb reconstruction with independent measurements. 
For all dual-comb measurements, the output beam from the chip is split using a beamsplitter.
A small fraction of the power is directed to a reference detector, while the remaining beam is transmitted through the sample arm. 
The reference and sample interferograms are recorded simultaneously, allowing common-mode timing and phase fluctuations to be corrected during post-processing.

The dual-comb signals are detected using a VIGO FIP-1k-1G-D photovoltaic mercury cadmium telluride detector module equipped with a PVI-4TE-10.6-1$\times$1 detector element. 
The specified 3~dB electrical bandwidth of the detector is 900~MHz, although the detector retains sufficient sensitivity to measure multiheterodyne signals beyond this nominal bandwidth. 
Time-domain traces are acquired using a fast oscilloscope (Tektronix MSO 71604C) with acquisition windows ranging from 5~\textmu s to 500~\textmu s. 
The repetition-rate detuning, $\Delta$, is set by the two RF injection frequencies and is locked by the RF sources; in the present experiments, $\Delta$ is tuned between 8~MHz and 11~MHz. 
This range allows the multiheterodyne spectrum to be placed within the available detector and digitizer bandwidth while maintaining short interferogram periods.

The acquired interferograms are corrected for fluctuations in the relative carrier-envelope-offset frequency, $\Delta f_{\mathrm{CEO}}$. 
Following an approach similar to that used in coherent correction of free-running QCL dual-comb signals~\cite{Sterczewski2019_2}, we track the phase evolution of the strongest RF comb tooth using the Hilbert-transform of the signal. 
The extracted instantaneous frequency provides a time-dependent estimate of the relative CEO drift between the two combs. 
Each interferogram is then phase-shifted onto a CEO drift-stabilized grid before Fourier transformation. 
This correction narrows the RF comb teeth and enables coherent averaging over many interferograms.

For continuously swept spectral measurements, the DC currents and RF injection frequencies are swept together so that the optical comb teeth move across the spectral gaps between adjacent longitudinal modes. 
At each sweep point, the dual-comb spectrum is reconstructed on the CEO drift-stabilized RF grid. 
The reproducibility of the spectral tuning algorithm allows successive spectra to be stitched together without an optical reference during each scan. 
An independently measured FTIR spectrum is used only to calibrate the absolute optical-frequency axis.
Adjacent spectra are compared with the preceding acquisition to verify the correct relative offset and to ensure continuous stitching across the full scan. 
The final stitched spectrum is obtained by combining all sweep points onto a common optical-frequency grid.

Gas sensing measurements are performed using a single-pass gas cell filled with 1.25\% nitrous oxide (N$_2$O) in air at atmospheric pressure. 
The optical path length through the gas is approximately 13~cm. 
The cell is terminated with ZnSe windows, each 8~mm thick, to provide high transmission near 8~\textmu m. 
For the sample measurement, the main dual-comb beam is transmitted through the gas cell before detection, while the reference arm bypasses the cell and is used for phase correction and normalization. 
The reconstructed transmission spectra are compared with HITRAN-based simulations using fitted values of the concentration, pressure, and path length.

\section*{Data availability}

Data underlying the results presented in this paper are not publicly available at this time but may be obtained from the authors upon reasonable request.

\section*{Author contributions}

T.P.L., J.F., and D.K. conceived the idea. 
J.F. and T.P.L. designed the devices, and J.F. created the mask and fabricated the devices.
J.F. and T.P.L. performed device characterization.
T.P.L., P.C., and M.O. performed dual-comb spectroscopy, and M.O., T.P.L., J.F., and D.K. analyzed the dual-comb data.
T.P.L. and J.F. wrote the manuscript with input from all authors.
B.S. and F.C. supervised the research.

\section*{Competing interests}
The authors declare no conflict of interest.

\begin{acknowledgments}
T. P. Letsou thanks David P. Burghoff (University of Texas at Austin) for valuable discussions on dual-comb signal reconstruction.
This material is based on work supported by the National Science Foundation under Grant No. ECCS-2221715.
J. Fuchsberger and B. Schwarz received funding from the European Innovation Council (EIC) Pathfinder project UNISON [101128598].
The authors gratefully acknowledge the Center for Micro- and Nanostructures (ZMNS) of TU Wien for providing the cleanroom facilities.
M.O. acknowledges funding from the European Union (grant agreement 101076933 EUVORAM). The views and opinions expressed are, however, those of the author(s) only and do not necessarily reflect those of the European Union or the European Research Council Executive Agency. Neither the European Union nor the granting authority can be held responsible for them. 
The authors acknowledge TU Wien Bibliothek for financial support through its Open Access Funding Programme.
\end{acknowledgments}

\newpage
\bibliography{dualcomb}
\newpage

\end{document}